# How do we assess computation in physics?

Hannah C. Sabo[1], Tor Ole B. Odden[1], and Marcos D. Caballero[1,2]

[1]Center for Computing in Science Education, University of Oslo, 0316 Oslo, Norway 2
[2]Department of Physics and Astronomy & CREATE for STEM Institute, Michigan State University, East Lansing, 48824 Michigan, USA

## Computation in physics education

In recent years, computing has become an important part of the way we teach and learn physics. Teachers, both at high school and college levels, now use computational activities in many of their courses[1]. Physics departments are offering specialized courses and degrees in computational physics. And many countries are adding programming and/or computational thinking to their secondary science education standards[2–4]. Although we know more about how to teach computation[5], that's only half the picture: we need to know how to assess it. In this paper, we provide a snapshot of some commonly-used assessment activities and forms.

When we use the term assessment *activity*, we are referring to a broad category of task, such as exams or assignments. When we use the term assessment *form*, we are referring to the focus and manner of assessment, such as simply checking the output or using a rubric.

## Assessment of computational physics is different from standard physics

Computational physics is pedagogically and conceptually different from standard physics; these differences are reflected in the distinct ways we teach and assess them. In standard physics (both traditional and reformed pedagogies), homework assignments and exams require students to analytically solve more complex problems, involving diagrams, derivations, or critical evaluation of solutions. Sometimes the result is an equation, other times it is simply a number.

In computational physics, we don't produce equations: rather, our goal is to produce and interpret models and data. Computational physics students produce computational artifacts, like simulations or data analyses written in scripts, and then use them to produce and interpret an output like a visualization. Despite these differences, computational artifacts and analytic equations both serve as approximations/models of physical phenomena.

The tools and resources for producing these models are different. Analytical physics uses standard physics tools like conservation laws, Newton's laws, etc., which are embedded in mathematical equations. We combine and manipulate these equations using mathematical tools like calculus and vector math, and thereby create a mathematical model of the phenomenon which describes its behavior in time or space. Computational physics also uses these same intellectual tools, but instead of mathematics we embed these tools in code and manipulate that code using computational syntax and standard algorithms, like numerical integration and grid techniques.

Given these differences, we, as instructors, have to think about computational physics assessment differently; it's not enough to simply tweak our standard assessments by adding a computational "flavor" to our existing assignments or exams, since this ignores the key differences between the two types of physics. In this regard, computational physics is somewhat similar to experimental physics; it has long been understood that assessing students' experimental physics skills and techniques is a different beast than standard theoretical physics[6].

For instance, computational physics problems are more modular than standard physics problems; computational problems have many different parts and dimensions that can easily be swapped in and out, depending on the problem. We're asking students to tell a computer to do a calculation and manage the output. Furthermore, with similar inputs and systems, we can ask the computers to do different things. So, as instructors, we have to make choices on which parts you are assessing - outputs, script, model, interpretation. The increased complexity also means that thorough assessments will likely take more time than a standard in-class physics assessment.

**Some forms and activities of computational physics assessment**

As a first step in this direction, we present a snapshot of current computational physics assessment *activities* and *forms*, based on common practice and published examples. In Table I, we summarize assessment activities, which are the types of tasks: e.g, exams, assignments. In Table II, we discuss assessment forms, which describe the foci and manner of assessment.

Our insight is that these two aspects of assessment are related but somewhat independent of one another; they can be combined in different ways. For example, a computational problem on a written exam (activity) could be assessed with a rubric or progress on steps (form). Research on computational thinking recommends using multiple assessment methods[7–9]; this gives students varied opportunities to express their understanding of computation in different ways.

Instructors can mix and match different types of activities and forms for better coverage of their students' understanding. Multiple-choice questions, an activity useful for quickly gauging students' conceptual understanding[10,11], can be paired with long answer questions, such as Parson's Problems or a debugging task, which can test both physics and programming knowledge. With take home exams, students can engage with longer, authentic problems. Similarly, different assessment forms yield different information about their students' understanding. For example, an instructor can quickly glean whether a code is functioning as expected by checking the outputs, while looking at students' interpretations takes more time, but yields insight into how students extract physical meaning from their computations.

Some types of activities, especially projects, allow students to deeply dive into authentic computational physics topics. Depending on the project type, there is a tradeoff between guidance and freedom. Closed-ended projects provide scaffolding and structure. Instructors can anticipate and troubleshoot places where students encounter difficulties. Whereas, open-ended projects foster student inquiry and ownership, but these projects might require more scaffolding for less experienced students. As longer projects progress, the assessment form changes. While students are working on the project, instructors can provide formative feedback by using assessment forms like monitoring students' reports, their progress on steps, and self-assessment. After submission, instructors can give summative assessment via rubrics, reports, and a variety of other forms.

Computational assignments are activities which can be used in class, in labs, or as homework to help students familiarize themselves with shorter computational techniques. For both closed-ended projects and assignments, instructors might use a quicker form of assessment, like checking a students' outputs or visualizations. If the outputs do not match expectations, instructors can look through the students' code and assign points based on progress on steps. If instructors sense programming anxiety, they can use a form which reduces anxiety, like participation-based.

**Parting thoughts:**

These assessment activities and forms are based on the last two decades of work. As Generative AI, such as ChatGPT, become increasingly common, they will likely change how we think of computational physics assessment. Currently, generative AI can produce code (though not necessarily "good" code—or physics!), and this will necessarily impact which assessment activities and forms we use.

We also see opportunities for importing ideas and philosophies from computational physics back into assessment of standard, analytical physics. For instance, assessment on a standard physics assignment or exam frequently focuses on catching students' errors, e.g., dropping a negative or using an inapplicable equation. What if, instead of framing these errors as mistakes, we instead considered them "bugs"—a natural, even welcome, part of any computational physics endeavor? Admittedly, in computational physics these bugs are easier to catch, since they are often reflected in error messages or strange outputs. However, if we frame errors as a natural part of the physics learning process, rather than problems or sources of judgment, it might help reduce some of the widespread feelings of intimidation that students often feel when taking physics. It might also open discussions of how to identify these "bugs", or even lead students to make sense of their physical significance. In this way computation could perhaps help foster a more resilient and empowered generation of physicists.

**Table I: Computational Physics Assessment Activities**

| Assessment Activity | Form | Example assessment tasks |
|---|---|---|
| **Exam (Summative)** | Written | Interpret/correct errors in a script |
| | | Parson's problems[12]: organize scrambled code lines |
| | | Fill in the missing lines of a pre-written script. |
| | Oral presentation | Present a project in front of the class and answer questions about it |

| | Multiple-choice | Given student an incorrect output, what caused the error |
| --- | --- | --- |
| | | Given a part of a script, choose the next lines of code to make the script work accurately |
| | | Identify which of the following is NOT a correct expression for [quantity of interest]? |
| **Take-home exam (Summative)** | Written | Choose-your-own-adventure problem: investigate a system with options to explore and several possible modifications |
| | | Use a familiar script to conduct a straightforward analysis |
| | | E.g. predict motion of an object using numerical integration |
| | | Write a script to produce an analysis |
| **Clicker questions (Formative)** | Lecture | Similar to multiple-choice examples above |
| **Open-ended project (Summative)** | Written | Computational essay[13]: combine prose, code, mathematics, and visualizations to explore/explain a concept or analysis |
| | | Compare a simulation, experiment, and/or analytic models of the same phenomenon[14] |
| **Closed-ended project (summative)** | | Follow steps to reproduce a computational physics result[15] |
| **Assignment** | Modeling | Analyze a system using a technique, like numerical integration, Monte-Carlo simulation, etc  E.g. Simulate the Ising model using a Monte Carlo |
| | | Modify an existing model.  E.g. add air resistance |

|  | Data analysis | Given a dataset, perform standard analyses like numerical integration or Fourier transforms, and interpret the results |
|  |  | Collect/find a dataset, then perform standard analyses (described above). |

Table II: Forms of Computational Physics Assessment

| Assessment Form | What is being assessed? | How do we use it? |
| --- | --- | --- |
| Output | Code/model produces correct output (defined in advance) | Quickly compare output to expectations; try program with different initial conditions |
| Progress on steps | Whether/how far the student progressed through the steps of a scaffolded assignment/project | Evaluate the overall progress and/or how well they did each step |
| Communication and interpretation | How clearly the student interprets and discusses the output of the computer into physical meaning | Pairs well with oral and written reports/projects. How much did they accomplish, and how well do they explain/present it? |
| Code quality and correctness | How the code compares to expectations of well-structured, documented, or efficient code | Either on its own or with another form. |
| Participation/ effort | Student participation | When introducing computation into physics classes to get students to engage with computation, in a low-risk setting. Based on good-faith effort. |
| Rubric[16] | Multiple categories:<br>● Code quality<br>● Model quality<br>● Scientific practices<br>● Physics use | Clearly define rubric categories and point allotment. Distribute the rubric to students and follow the scoring guidelines |
| | • Communication (Introduction, conclusion, writing, documentation) | |
| Self-/peer assessment | Students' evaluations of their own/classmates' work (likely using rubric) | Pair with a rubric, model appropriate/constructive criticism throughout the project, invite students to reflect on their own progress |